# Machine Learning-based Efficient Ventricular Tachycardia Detection Model of ECG Signal

[1]Pampa Howladar, [2]Manodipan Sahoo
[1]Indian Institute of Engineering Science and Technology, Shibpur, India,
[2]IIT (ISM), Dhanbad, India
p.howladar.@gmail.com,  manodipan@iitism.ac.in

**ABSTRACT**

In primary diagnosis and analysis of heart defects, an ECG signal plays a significant role. This paper presents a model for the prediction of ventricular tachycardia arrhythmia using noise filtering, a unique set of ECG features, and a machine learning-based classifier model. Before signal feature extraction, we detrend and denoise the signal to eliminate the noise for detecting features properly. After that necessary features have been extracted and necessary parameters related to these features are measured. Using these parameters, we prepared one efficient multiclass classifier model using a machine learning approach that can classify different types of ventricular tachycardia arrhythmias efficiently. Our results indicate that Logistic regression and Decision tree-based models are the most efficient machine learning models for detecting ventricular tachycardia arrhythmia. In order to diagnose heart diseases and find care for a patient, an early, reliable diagnosis of different types of arrhythmia is necessary. By implementing our proposed method, this work deals with the problem of reducing the misclassification of the critical signal related to ventricular tachycardia very efficiently. Experimental findings demonstrate satisfactory enhancements and demonstrate high resilience to the algorithm that we have proposed. With this assistance, doctors can assess this type of arrhythmia of a patient early and take the right decision at the proper time.

*Keywords:*
*Electrocardiography (ECG), ECG signals, filtering, data classification, feature extraction, ventricular tachycardia arrhythmia*

## 1. INTRODUCTION

The risk of sudden death may increase with ventricular tachycardia (VT), which is normal in patients with structural heart disease. Most ventricular arrhythmias are triggered by cardiomyopathy, or hypertension, ST-segment changes, chronic obstructive pulmonary disease (COPD), and instant death happens unless correctly diagnosed or treated [1]. Sudden death constitutes more than one-half of all heart deaths and up to 15% of the overall United States mortality. Ventricular arrhythmia is characterized by an irregular ECG rhythm which accounts for 75%–85% of sudden death in people with cardiac defects unless it is treated within seconds. In infants, VT is rare but can be present in the presence of structural heart disease. Overall, VT is frequent in men rather than women [2].

Over the last few decades, clinical, surgical, and technical advances have seen an extraordinary pace. Since then, significant efforts were made to benefit from advancements in technologies and the use of computers in the medical field. As the electrocardiogram (ECG) is used to analyze the most significant organ of the human body, the heart's physiological symptoms, ECG analysis with maximum accuracy is of considerable interest to cardiologists [3]. There has been a lot of research in the field of ECG analysis aimed at automatic signal detection with almost perfect rates. Efforts have been focused over the past thirty years to model the skills of cardiologists and specialists via computers. Many researchers [4-8] have addressed this issue by developing a variety of alternative ECG identification and QRS detection algorithms that have become widely acknowledged in the literature. Among the various methods examined and assessed, the machine learning approach took particular attention due to their properties, such as nonlinearity, learning capability, and universal approach, which permit them to solve challenges in the detection of distinctive features in ECG signals such as QRS detection and VT diagnostics. In this paper, we will discuss our approach to design and implement an improved machine learning model for the detection of ventricular tachycardia.

### 1.1 Electrocardiogram Signals:

The electrocardiogram (ECG) is a record of the electrical activity produced by the heart on the body surface. During ECG, our chest and often our limbs are connected to sensors (electrodes), which are able to sense the electrical activity of the heart. An ECG tracks the timing and duration of each heartbeat's electrical phase.

In 1899, Waller first noticed the ECG. In 1903, Einthoven named the waves of the ECG with 398 electrophysiological principles still in use today. He attributed the P-to-U letters to waves, which prevented contradictions with other physiological waves that he studied. A typical ECG signal is seen in Figure 1. The ECG signals usually range between ±2 mV and 0.05–150 Hz bandwidth [9]. The morphology of ECG waves is based on how much tissue is activated for a time unit and on the relative cardiac activation rate and direction.

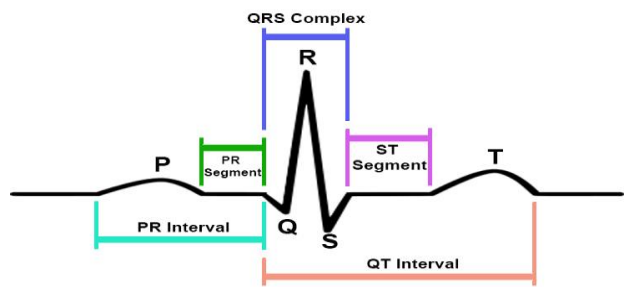

Fig1: A diagram of a normal ECG signal

As a result, the physiological potentials of the pacemaker, i.e. the sinoatrial (SA) node, produced by a relatively small myocardial mass are not detectable on ECG. The P-wave representing atria depolarization is the first ECG wave in the cardiac cycle. The cardiac impulse travels from the atria to the ventricles via a series of specialized cardiac structures (the AV node and the His-Purkinje system). The small isoelectric section after the P wave is the PQ interval, which is attributed to the propagation delay (0.2s) triggered by the AV (atrioventricular) node. Once a ventricle's broad muscle mass is excited, the ECG surface is quickly and thoroughly deflected. The QRS complex or R Wave reflects the depolarization of the ventricles. A second isoelectric section, the ST interval, is observed following the QRS complex. The ST interval indicates the depolarizing time after activation of ventricular cells, usually varying from 0.25s to 0.35s. Following the conclusion of the ST section, the ventricular cells return to their electrical and mechanical state of rest and conclude the repolarization cycle, observed as the T wave signal of the low frequency. In certain individuals, at the end or after the T wave there is a small peak called the U wave. Its origin was never completely known but is thought to be a potential for repolarization [7].

**1.2 Ventricular Tachycardia Arrhythmias**

Arrhythmias are attributed to cardiac disorders that make heartbeats erratic. Heart Rhythm problems (heart arrhythmias) arise when your heartbeats are not coordinated properly by electrical pulses, which make your heart beat too quickly, too slowly, or irregularly. Cardiac arrhythmias can be categorized as either ventricular arrhythmia (the ventricle origin) or supraventricular arrhythmias according to their origins (originate in parts of the heart above the ventricles, usually the atria). It may also be classified based on how they affect the heart rate. Bradycardia is identified when the heart rate is less than 60 beats per minute, while tachycardia is diagnosed when the heart rate is more than 100 beats per minute at the cardiac stage.

There are common types of cardiac arrhythmias. VT is one of the major cardiac arrhythmias among them. It could last for several seconds (sustained VT) or several minutes or even several hours (non-sustained VT). Sustained VT is a risky rhythm that sometimes leads to ventricular fibrillation if it is not treated. This ventricular fibrillation is too dangerous as here ventricles quiver ineffectively without a real heartbeat resulting in brain damage, unconsciousness, and death within minutes.

Ventricular tachycardia is an irregular cardiac rhythm that starts in the right and left ventricles. Non-sustained VT is considered to be large-complex tachyarrhythmia (QRS length longer than 120 milliseconds) at a heart rate of more than 100 beats a minute.

*Symptoms:* No symptoms or mild chest fluttering may result from Non-sustained VT. Sustained VT is typically deadly and causes lightheadedness or loss of consciousness.

*Treatment*: If no structural damage happens to the heart, non-sustained VT does not require to be treated. In sustained VT, either with an intravenous injection or electrical emergency shock (defibrillation) is needed to regain the heart's normal rhythm

**2. RELATED WORKS AND OUR CONTRIBUTION**

**2.1 Related Works:** Ventricular arrhythmias have been identified in a variety of techniques based on morphological, spectral, and mathematical properties taken from the ECG signal [10-13]. A useful method for enhancing detection performance was also proposed for machine learning technologies such as neural networks [14] and Supervised Vector Machine [15]. Although these techniques have shown advantages in ventricular arrhythmia diagnosis, they have some limitations. Some techniques are too complex to enforce or compute, some of them have no precision to distinguish between regular and irregular circumstances and all of them retain the period of late diagnosis, typically not sufficient for intervention.

**2.2 Our Contributions:**

In this paper, we propose a cardiac arrhythmia prediction model for ventricular arrhythmia detection. The major contributions of this paper are as follows:

- In the pre-processing stage, we detrend and denoise the signal to eliminate the noise for detecting features properly.

- After that, the QRS complex having high amplitude and its R-peak detection has been performed using our proposed technique.
- Once this R-peak has been detected, its related parameters like RR interval, QRS duration, and HBR (Heartbeat rate) are calculated.
- Using these parameters, an efficient multiclass classifier-based machine learning-based model has been developed to classify various types of ventricular tachycardia arrhythmias.

As far as we are aware, it is the first that various machine learning-based classification techniques are analyzed and then select one technique according to their high performance in order to detect ventricular tachycardia arrhythmias of ECG signal.

The remaining part of this paper is structured accordingly. In Section 3, material and method have been presented. Section 4 presents the result and discussion. Finally, concluding remarks are presented in Section 5.

## 3. MATERIAL AND METHOD

### 3.1 Material

A number of ECG databases are available in PhysioNet. For our paper, ECG data are collected from MIT-BIH svdb, Cudb, MIT-BIH nsrdb, and MIT-BIH Arrhythmia Database (mitdb) of PhysioNet database [16] and analyzed to find any differences in the ECG signal characteristics. Each record of MIT-BIH svdb, cudb, MIT-BIH nsrdb and mitdb datasets includes 230400 samples, 127,232 samples, 11730944 samples, and 650000 samples respectively. These sets reflect various subject categories and conditions for recording, such as sampling rates (128 Hz, 250 Hz, and 360 Hz) and interferences. ECG1 data are used without exclusion for any record. By extracting information about Heartbeat rate (HBR), R-R intervals, QRS amplitudes, the onset of the ventricular arrhythmia can be detected.

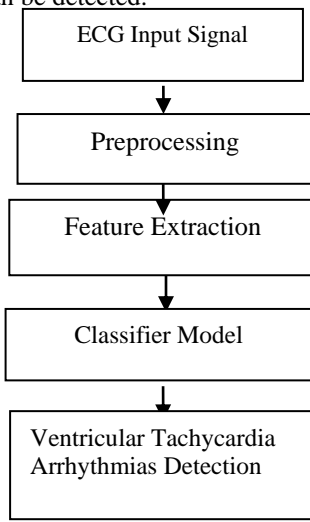

Fig 2: A block diagram of proposed method's stage

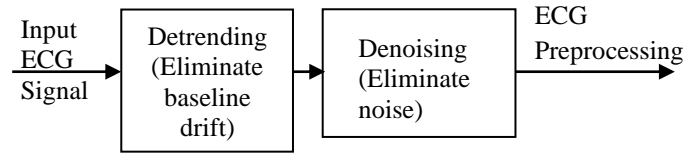

Fig 3: ECG Signal Preprocessing before feature extraction

### 3.2 Method:

Our proposed method is composed of the following steps (Fig 2.)

### 3.2.1 Data preprocessing:

In electrocardiograms, a variety of noises can also be recorded from very high and low frequency [7] which induces baseline drifts and signal noise in the ECG and is very difficult to diagnose clinically. Noise from the signal must be eliminated for proper ECG diagnosis.

Fig 3 shows the necessary possible steps of ECG signal preprocessing. A method of eliminating the baseline signal drift [07, 08] is called as a detrending, and the signal noise removal process is referred to as denoising. These two signal processing techniques fall under the domain of the ECG signal preprocessing.

In order to eliminate noise artifacts, an ECG signal must be preprocessed. Noise from multiple sources such as breathing, electrically charged electrodes, or moving subjects may distort the ECG signal. This low-frequency noise also needs to be considered as that can impact the peak detection mechanism. The R peak detection may not be influenced by noise because the peak amplitude is high. However, since having low amplitude the P, Q, S, and T wave detection are influenced by the noise. The lower frequency components in the ECG signal, according to the American health association (AHA), are approximately 0.05 Hz [17]. So, frequencies of 0- 0.05 Hz are essential to remove to lower the baseline drift. Therefore, band-pass filtering is a required first step for nearly all QRS detection algorithms.

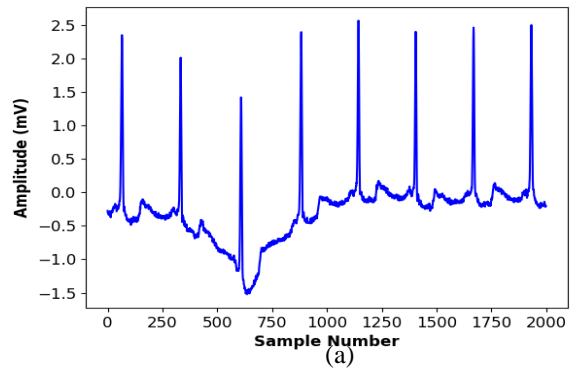

(a)

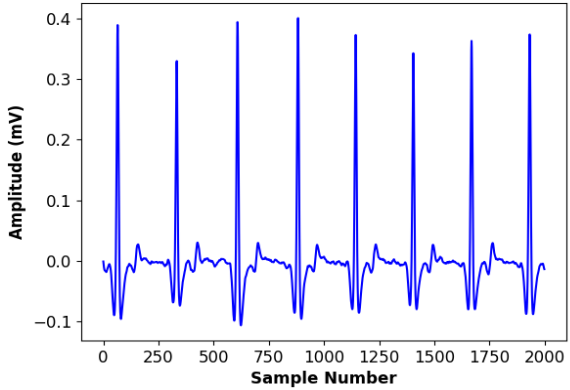

(b)

Fig 4: (a) Original signal cu01.png (b) detrending (eliminate baseline signal drift of cu01)

The goal of applying bandpass filtering is to remove the baseline drift and high frequency that does not detect QRS complexes. This baseline drift and unnecessary high frequencies can be eliminated with a bidirectional butterworth bandpass filter [07, 08]. Once the signal has been preprocessed, it can be used for further processing. In Fig. 4 baseline drift has been eliminated on record cu01 and in Fig 5 noise has been eliminated on record cu14.

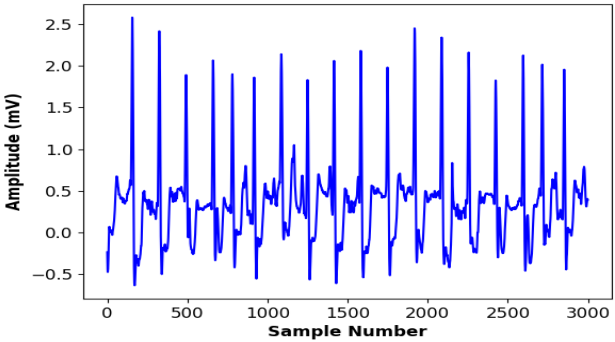

(a)

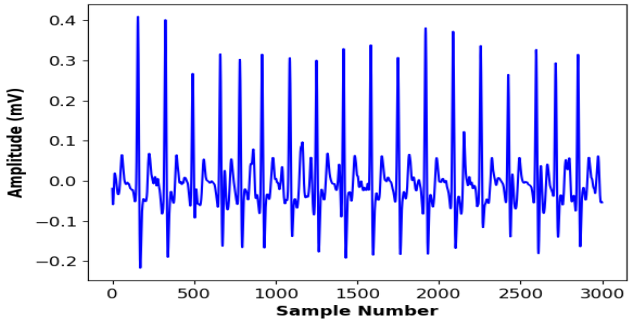

(b)

Fig5. (a) Original signal cu14.png (b) denoising (eliminate noise)

**3.2.2 Feature extraction:** After preprocessing completed required features (Q, R and S) are extracted from the noise free signal.

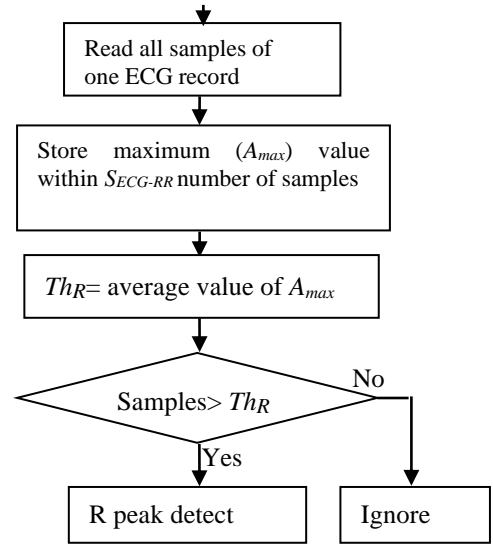

Fig 6: Proposed R peak detection method

In order to detect the QRS complex, we have proposed the peak detection technique, as shown in Fig 6. First, the method reads all samples of one ECG record that is obtained from SRAM in which all required samples are stored for this work. Now, the total number of samples ($S_{ECG}$) and total time ($t_{ECG}$) with respect of the total number of samples are recorded. So, time taken for each sample is $t_{ECG-s} = t_{ECG} / S_{ECG}$.

Now we know from Table I that the normal range of RR interval is (0.6 -1) s. Let's say we assumed that $t_{RR}$ is having a range $0.6 < t_{RR} < 1$. Therefore total number of samples between RR intervals for normal ECG signal must be $S_{ECG-RR} = t_{RR} / t_{ECG-s}$.

After that, peak values are found with $S_{ECG-RR}$ number of samples and are stored in an array. Once we stored all values, the average value is to be calculated. This value is considered as threshold value $Th_R$. Now, every sample value of provided ECG signal is compared with $Th_R$. if the sample value of the ECG record is greater than $Th_R$ then that value is considered as R-peak value. In this way, R-peak has been detected. Finally, all required parameters are calculated from this feature as shown in Figs. 7 and 8.

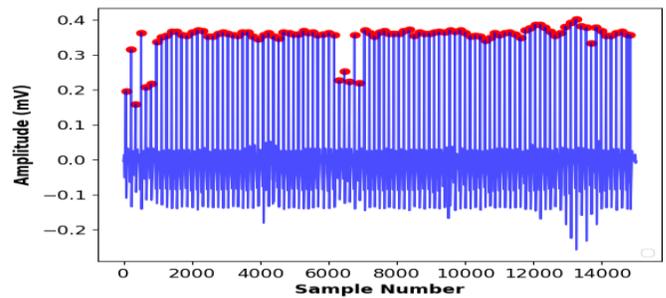

Fig 7: R peak detection and HBR calculation of Cu13 record

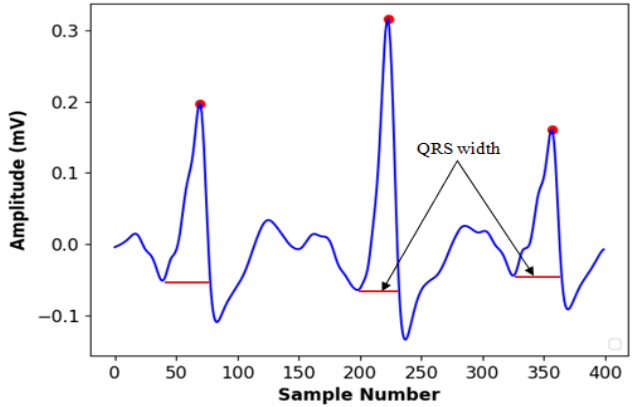

Fig 8: QRS width is marked on ECG Signal

After that necessary parameters using these features are calculated. These values are compared with the values of the standard ECG signals to determine the criticality of the signal. For ventricular tachycardia detection using machine learning the various parameters considered are:

- RR interval

- Heartbeat rate (HBR)

- QRS duration

Table I [8] presents essential characteristics of the ECG signal and its normal range.

TABLE I. NORMAL RANGE OF REQUIRED ECG SIGNAL

| PARAMETERS | NORMAL RANGE |
|---|---|
| **RR interval** | 0.6-1s |
| **QRS duration** | Upto 0.10s |
| **HBR** | 60-100 BPM |

**3.2.3 Classifier model:** Here we concentrate on the proper classification of the ECG signal which will lead to the creation of a ventricular tachycardia arrhythmias detection-based system with machine learning techniques. The signal beats obtained from ECG are divided into two distinct categories: critical and non-critical. The primary goal is to accurately identify the ECG sample beat and to minimize errors. Therefore, it is very important not to misclassify any critical signals as non-critical that can cause severe problems in the system. These issues are already discussed in the paper.

All the parameters (RR interval, QRS duration, HBR) obtained after feature extraction are used as inputs for our classifier model. Precision, sensitivity or recall, and F1 score are used in order to assess the efficacy of applied machine learning such as KNN, SVM, logistic regression, decision tree whereas precision is the percentage of signal that is already labeled as a critical which is potentially critical. Sensitivity or recall is the percentage of the critical signal that is marked as critical. The F1 score or specificity is the harmonic mean of precision and sensitivity, typically more useful than accuracy when a class distribution is unequal.

In order to classify essential characteristics that influence the model, the weight matrix of neural networks is also studied. These three measurements (precision, sensitivity, and F1 score) originate from the confusion matrix and are compared by F1 scores. We choose our classified model best on their performance.

## 4. RESULT AND DISCUSSION

### 4.1 Method of the presence of noise detection and the filtering effect of ECG signal:

Fig 9 explains how to detect the presence of noise in a signal whereas for ventricular tachycardia 110 bpm <$HBR_i$, $HBR_f$ < 250 bpm. Fig 10 shows two identical signals with no filter and applied filter. From these two figures, it is easily visible that the criticality of some signals is possible to remove only after noise remove. Fig 10(a) shows a noisy signal having baseline drift and noise. This baseline drift and noise has been removed using butterworth filter, shown in fig 10 (b).

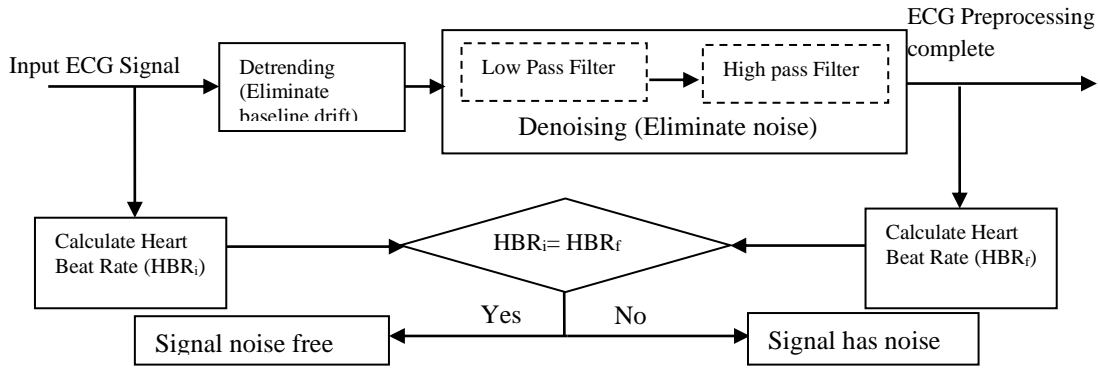

Fig 9: Signal preprocessing after feature extraction

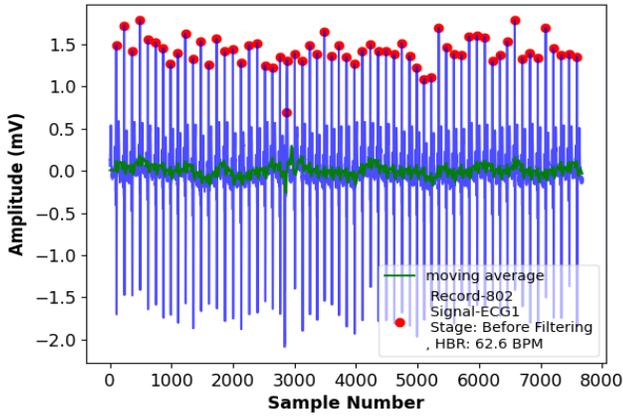

(a)

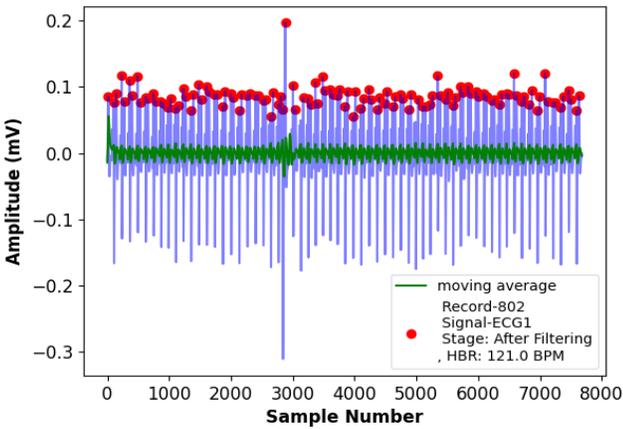

(b)

Fig 10: (a) Heart Beat Rate calculation before filtering (b) Heart Beat Rate calculation after filtering

### 4.2 Simulation results:

For ECG signal filtering, butterworth filter has been utilized in Python 3.7, and also for training and testing, Python 3.7 is used in all simulations. The characteristics of the various signals found in these studies show with a small number of signals in Table II.

Abnormalities observed of these ECG signals after filtering are shown in this table.

Our proposed algorithm has been assessed by measuring the sensitivity, precision, F1 score and accuracy.

These are the following definitions:
1) The precision (*PR*): *PR=TP / (TP+FP)*
2) The sensitivity (*SE*): *SE= TP/ (TP+FN)*
3) The specificity (*SP*): *SP=TN/ (TN+FP)*

Where: *FP = False Positives; TP = True Positives; FN = False Negatives; TN = True Negatives;* and
*N= FP + FN +TP +TN.*

From Table III, while considering accuracy data it is observed that logistic regression has outperformed KNN, SVM, and MLP (neural network). Logistic regression and decision tree perform equally better. While considering the F1 score, here also it is observed that both logistic regression and decision tree have the best F1 score, and then comes MLP (neural network), SVM, and KNN in descending order. Therefore for detecting critical ECG signal bits, logistic regression and decision tree are preferable. Table IV shows multi-class classification among non-critical, sustained VT, and non-sustained VT signals of our logistic regression or decision-based model.

In Table V, our experimental results are compared with [18-20]. Experimental findings demonstrate satisfactory enhancements and demonstrate high resilience to the algorithm that we have proposed.

TABLE II. RESULTS OF DIFFERENT EXTRACTED FEATURES WITH SMALL NUMBER OF SIGNALS

| Features | Normal range | Record cu08 | | Record cu12 | | Record cu32 | | Record cu34 | |
|---|---|---|---|---|---|---|---|---|---|
| | | Before filter | After filter | Before filter | After filter | Before filter | After filter | Before filter | After filter |
| RR interval | 0.6-1s | >1 | <0.6 | 0.67 | <0.60 | 0.60 | 0.29 | 0.74 | <0.6 |
| QRS width | Upto 0.10 sec | <0.10 | >0.12 | <0.10 | <0.10 | >0.10 | <0.10 | <0.10 | >0.12 |
| HBR | 60-100 BPM | 46 | 169 | 100 | 100 | 108 | 98 | 81 | 140 |
| Ventricular Tachycardia | | no | yes | No | yes | no | yes | no | yes |

TABLE III. PERFORMANCE ANALYSIS

| Algorithm | Precision | Sensitivity | F1 Score | Accuracy |
|---|---|---|---|---|
| KNN | 0.93 | 0.92 | 0.92 | 0.91 |
| SVM | 0.95 | 0.95 | 0.94 | 0.95 |
| Logistic Regression | 0.98 | 0.97 | 0.97 | 0.97 |
| Decision Tree | 0.97 | 0.97 | 0.97 | 0.97 |
| MLP(Neural Network) | 0.95 | 0.95 | 0.95 | 0.95 |

TABLE IV. MULTICLASS CLASSIFICATION FOR LOGISTIC REGRESSION AND DECISION TREE BASED MODEL

| Algorithm | Precision | Sensitivity | F1 Score |
|---|---|---|---|
| Non VT | 0.89 | 1.00 | 0.94 |
| Sustained VT | 1.00 | 0.91 | 0.95 |
| Non-sustained VT | 1.00 | 1.00 | 0.97 |
| Accuracy | | | 0.97 |
| Macro avg | 0.96 | 0.97 | 0.97 |
| Weighted avg | 0.97 | 0.97 | 0.97 |

TABLE V. COMPARATIVE STUDY

| Author | Precision | Recall | F1-score | Accuracy |
|---|---|---|---|---|
| Zihlmann et al. [18] | - | - | 79% | 82% |
| Goodfellow et al. [19] | 84% | 85% | 85% | 88% |
| Jalali et al.[20] | 86% | 86% | 85% | 89% |
| Proposed method | 98% | 97% | 97% | 97% |

## 5. CONCLUSION

In this paper, we proposed a machine learning-based method in order to detect ventricular tachycardia-based arrhythmia of a signal. Our proposed method follows preprocessing stage, feature extraction, and classifier model. In the pre-processing stage, we detrended and denoised the signal in order to eliminate the noise for detecting features properly. After that, the QRS complex having high amplitude and its R-peak detection has been performed using our proposed technique. Once this R-peak has been detected, its related parameters like RR interval, QRS duration, and HBR are calculated.

Using these parameters, we prepared one efficient machine learning-based multiclass classifier model which is able to classify various types of ventricular tachycardia arrhythmias with very high accuracy. This work deals with the problem of reducing the misclassification of the critical signal very efficiently. To our best knowledge, it is the very first to evaluate different methods based on machine learning and then pick a suitable method based on its high level of efficiency in order to detect ventricular tachycardia arrhythmias of ECG signal. Experimental results indicate satisfactory developments in our proposed algorithm and appear to be highly stable.

Since perfection in the medical sector is necessary, the F1 score to identify critical signals must be further enhanced and this can be achieved if our model is exposed to a wider range of data. It is observed from our results that, Logistic regression and Decision tree-based models are the most efficient machine learning models for detecting ventricular tachycardia arrhythmia. We can train using more sophisticated algorithms to generate better results with more data. Also classifying critical signs in the less critical group may be a helpful approach for the doctor's support in making the right decision.

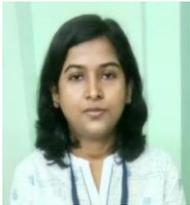

**Pampa Howladar** completed her B. Tech in Information technology from Govt. College of Engineering & Ceramic technology (GCECT) under MACAUT University. She received M.Tech degree in School of VLSI Technology from Indian Institute of Engineering Science and Technology, Shibpur (Formerly BESU) in in 2013. She received PhD degree in Information Technology in the area of Design and Optimization of Digital Microfluidic Biochips on Microelectrode-Dot-Array Architecture from Indian Institute of Engineering Science and Technology, Shibpur, India in 2020. Her research interests are digital microfluidic biochips, Embedded Systems, Bioinformatics and Biomedical Signal Processing. Her research works appeared in reputed journals including IEEE transactions on Very Large Scale Integration (VLSI) Systems and IEEE/ACM Transactions on Computational Biology and Bioinformatics and also in referred international conference proceedings. She is now actively exploring machine learning and artificial intelligence techniques for biomedical applications. She has got best paper awards for her papers in Computing, Communication and Sensor Network Conference in 2018, International Conference on Microelectronics Circuit &Systems in 2021 and International conference on Data Analytics &Management in 2021. She also served as a Program Committee member of IEEE ISDCS, 2020.

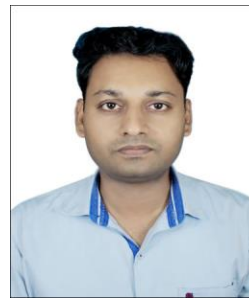

MANODIPAN SAHOO (M'14) was born in Haldia, West Bengal, India in 1983. He received M. Tech. in Instrument Technology from Indian Institute of Science, Bangalore in 2006. He received PhD degree from IIEST, Shibpur, India in 2016. His PhD thesis was on ``Modeling and Analysis of Carbon Nanotube and Graphene Nanoribbon based Interconnects". He is currently serving as an Assistant Professor in the Department of Electronics Engineering, Indian Institute of Technology (Indian School of Mines), Dhanbad, India. His research interests include Modeling and simulation of nano-interconnects and nano-devices, VLSI Circuits and Systems, Internet of Things, Biomedical Signal Processing . He has published more than 50 articles in archival journals and refereed conference proceedings. He has also published a Book entitled ``Modelling and Simulation of CNT and GNR Interconnects" with Lambert Academic Publishers in 2019. He published a book chapter entitled ``Modelling Interconnects for Future VLSI Circuit Applications" with IET Publishers in 2019. He is also associated as Member of IEEE, IETE, IEI and Life Member of Instrument Society of India.